\documentstyle[aps,prl,twocolumn,floats,epsfig]{revtex}
\begin{document}
\draft
\wideabs{
\title{Electron correlation in the Si(100) surface}
\author{\'Oscar Paz,$^1$ Ant\^onio J. R. da Silva,$^2$ 
Juan Jos\'e S\'aenz,$^{1,3}$ and Emilio Artacho$^{1,3}$}
\address{
$^1$Departamento de F\'{\i}sica de la Materia Condensada, C-III,
Universidad Aut\'onoma, 28049 Madrid, Spain \\
$^2$Instituto de F\'{\i}sica, Universidade de S\~ao Paulo, 
CP 66318, 05315-970 - S\~ao Paulo, SP  Brazil \\
$^3$Instituto Nicol\'as Cabrera, Universidad Aut\'onoma, 28049 Madrid, Spain 
}
\date{\today}
\maketitle
\begin{abstract}
Motivated by the controversy between quantum chemists and solid-state
physicists, and by recent experimental results, spin-polarized
density-functional (DFT) calculations are used to probe electron correlation
in the Si(100) reconstructed surface.
   The ground state displays antiferromagnetic spin polarization for
low dimer inclinations indicating, not magnetic order, but the 
importance of Mott-like correlations among dangling bonds. 
   The lowest energy corresponds to a higher dimer inclination with 
no spin. DFT energies, however, should be taken with caution here.
   Our results together with quantum-chemical findings suggest dimers 
with highly correlated electrons that tend to buckle due to interactions
with other dimers.
\end{abstract}
}

The freshly cut (100) surface of silicon has two dangling
bonds per Si atom. This is a highly unfavorable configuration,
and the surface reconstructs forming rows of dimers, thereby
reducing the number of dangling bonds by a half.
  This chemical description is clear and well established
\cite{review}, but the physics of the remaining two dangling
bonds per dimer is not. 
  The main reason is that the $\sigma$-like chemical bonds in 
silicon are well described by many electronic-structure methods, 
whereas the dangling bonds are much more subtle.
  The early argument for asymmetric dimers based on relating
symmetric dimers to metallic character \cite{Chadi79} was found to be
incorrect, since it is possible to obtain a symmetric, non-metallic
state within a Hubbard like framework \cite{Artacho89,Vinchon,Flores}.
  However, experimental evidence accumulated afterwards strongly 
suggesting the asymmetry of the dimers. 
  In addition, solid-state first-principles calculations based
on the local-density approximation (LDA) to density-functional
theory (DFT) converged to results in agreement with experiments
\cite{Ramstad}.
  The dimer asymmetry was rationalized in chemical terms, the
driving force staying within each dimer: a rehybridization of
the orbitals with a charge transfer from the lower to the upper
atom, leaving the lower with an $sp^2$-like hybridization that
would open the one-particle gap and thus lower the total energy
\cite{review,Pauling}. 

  The issue about the symmetry/asymmetry of the dimers would seem
to be settled and of no further relevance now, except for two
facts: ($i$) Kondo {\it et al.} \cite{Kondo} have recently 
performed low-temperature (20 K) scanning tunneling microscopy,
and they found that at these low temperatures the ``symmetric
dimers dominate the surface,'' with the asymmetric dimers being
dominant at higher temperatures (110 K). Based on these results
they claim that the ground state for the Si(100) system consists
of symmetric dimers, and that the theory should be revised in the
line of correlated electrons. It is worth mentioning that a similar
result was also obtained by another experimental group using the same
technique \cite{Yokoyama}, but they propose that the reason for the
observed symmetric appearance of the dimers is an anomalous flipping
motion of the buckled dimers; and ($ii$) different groups in quantum
chemistry \cite{Redondo,Carter,Paulus} have shown that a proper
treatment of electron correlations produces symmetric dimers with
a highly correlated ground state \cite{Artacho89,Vinchon}.

  From the point of view of the theory, the controversy can be
explained by noting that: ($i$) solid-state DFT calculations are
performed for realistic infinite-surface (slab) geometries, but
cannot describe the electronic correlations in a controlled way;
and ($ii$) quantum-chemistry calculations can describe electron
correlations properly, but have to work on isolated clusters
containing up to a few dimers.
  It is clear that a theory with appropriate geometry {\it and}
correlation description is needed. A theory that (we conjecture)
would show that if the dimers are asymmetric it is because of
reasons {\it extrinsic} to the dimer, i.e., interactions among 
dimers. The dimers would display important Mott-like
(antiferromagnetic) correlations, and there would be no 
driving force towards asymmetry within the dimer.

  Spin polarization is the simplest way to assess the importance
of Mott-like correlations. A spin-polarized wave-function tends
to the right dissociation limit, except for the fact that the
spin symmetry is artificially broken. 
  A spin-polarized DFT study of the surface is presented below. 
The polarization itself is to be interpreted as an indication
of the need of a better treatment of dynamical correlations,
and, therefore, the numbers are to be taken with caution. 
  The magnetic order is artificially imposed and of no relevance 
to the point, except as a way of describing the short range 
correlations.

  The DFT calculations were made using the numerical atomic-orbital 
(NAO) method \cite{pablo,daniel,emilio} in the {\sc Siesta} code 
implementation \cite{daniel}.
  The generalized gradient approximation (GGA) of Perdew, Burke,
and Ernzerhof~\cite{PBE} to Kohn-Sham~\cite{Kohn-Sham} theory was
the functional for exchange and correlation, except for the
LDA~\cite{PZ} tests shown below.
  Core electrons were replaced by norm-conserving
pseudopotentials~\cite{JLM} in their fully non-local
formulation~\cite{KB}.
  Valence electrons were described using a double-$\zeta$ polarised 
basis set of numerical atomic orbitals with a range defined by
an energy shift of 200~meV, and a variationally optimized 
splitnorm of 0.3~\cite{emilio}.
  A uniform real-space mesh with a plane-wave cutoff of 80~Ry was 
used for numerical integration.

  Calculations were performed for the 2$\times$1, $p$(2$\times$2)
and $c$(4$\times$2)
reconstructions of Si(100) using a repeated slab geometry
with nine layers of silicon and a layer of hydrogen atoms
saturating one of the sides of the slab.
  The H-Si bonds were relaxed in a bulk-like saturated slab
with H on both surfaces, using the bulk lattice parameter
obtained within the theoretical framework described above. 
  Then, one of the surfaces was kept 
saturated and fixed, including the H layer and the closest
Si layer. The remaining eight layers were allowed to
relax freely in all the calculations described below.
  Relaxations were considered to be finished when the maximum 
residual force was below 0.02~eV/\AA. The forces on the fixed
layers always stayed $\lesssim$0.06~eV/\AA.
  Integrations over the first Brillouin zone were approximated by
Monkhorst-Pack \cite{Monkhorst-Pack} sets of $k$-points for
a length cutoff of 15~\AA\ \cite{Moreno}, which corresponds
to a 4$\times$8$\times$1 set in the 2$\times$1 geometry.

  To test the methodology we have reproduced earlier 
results obtained with LDA. For bulk silicon a lattice
parameter of 5.40~\AA\ is obtained, which compares very well
with the experimental value of 5.43~\AA~\cite{Exp.values}
and with 5.37-5.38~\AA\ of LDA plane-wave (PW)
calculations~\cite{PW.values,javier}.
The bulk modulus is 95~GPa employing the Murnaghan
equation of state~\cite{murnaghan} and 96~GPa using a quartic 
fit, to be compared with 99~GPa of experiment~\cite{Exp.values}
and 96-98~GPa of PW~\cite{PW.values,javier}.
\begin{table}[t!]
\caption{LDA results. Dimer bond length ($d$), buckling angle
($\alpha$), energy difference with respect to symmetric 2$\times$1
relaxed configuration ($\Delta E$), displacement of subsurface atoms
in the direction parallel to the dimer rows ($\Delta x$), 
and surface energy ($E_s$)\protect\cite{surf. energy}.}
\begin{tabular}{lddddd}
Authors  &  $d$ & $\alpha$ & $\Delta E$ & $\Delta x$ & $E_s$ \\
 & (\AA) & & (eV/dimer) & (\AA) & eV/(1$\times$1) \\
\hline
DS\tablenote{J. D\c{a}browski and M. Scheffler, Ref. \cite{Dabrowski92}}
                &   &   &     &   & \\
~~s 2$\times$1  & - & - & 0.0 & - & - \\
~~a 2$\times$1  & - & 15$^{\rm o}$ & $-$0.1 & - & - \\
\hline
FP\tablenote{J. Fritsch and P. Pavone, Ref. \cite{Fritsch95}} & & & & & \\
~~s 2$\times$1     & 2.26 &  0.0$^{\rm o}$ &  0.000 &    -      & - \\
~~a 2$\times$1     & 2.28 & 16.9$^{\rm o}$ & $-$0.127 &    -      & - \\
~~$p$(2$\times$2)  & 2.33 & 18.5$^{\rm o}$ & $-$0.197 & $\sim$0.1 & - \\
~~$c$(4$\times$2)  & 2.32 & 18.2$^{\rm o}$ & $-$0.199 & $\sim$0.1 & - \\
\hline
KP\tablenote{P. Kr\"uger and J. Pollmann, Ref. \cite{Kruger93}} & & & & & \\
~~s 2$\times$1  & 2.25 &      -       &  0.00 & - & - \\
~~a 2$\times$1  & 2.25 & 19$^{\rm o}$ & $-$0.14 & - & - \\
\hline
GN\tablenote{A. Garc\'{\i}a and J. E. Northrup, Ref. \cite{alberto}}
                   & & & & & \\
~~a 2$\times$1     & - & - & - & - & 1.40 \\
~~$c$(4$\times$2)  & - & - & - & - & 1.36 \\
\hline
N\tablenote{J. E. Northrup, Ref. \cite{Northrup93}} & & & & & \\
~~$c$(4$\times$2)  & 2.29 & 17.8$^{\rm o}$ & $-$0.14 & 0.08 & 1.4 \\
\hline
This & & & & & \\
work & & & & & \\
~~s 2$\times$1     & 2.29 &  0.0$^{\rm o}$  &    0.00 & 0.00 & 1.57 \\
~~a 2$\times$1     & 2.28 & 17.5$^{\rm o}$  & $-$0.13 & 0.00 & 1.50 \\
~~$p$(2$\times$2)  & 2.34 & 18.3$^{\rm o}$  & $-$0.23 & 0.12 & 1.45 \\
~~$c$(4$\times$2)  & 2.34 & 18.5$^{\rm o}$  & $-$0.24 & 0.12 & 1.44 \\
\end{tabular}
\label{table:d2e}
\end{table}
  Table I shows the relevant magnitudes obtained for the 
2$\times$1, $p$(2$\times$2) and $c$(4$\times$2) reconstructions
as compared with the
figures obtained from LDA calculations of other groups.
Spin polarized results for LDA were never reported, probably
because the calculations converged to non-polarized. We
obtain non-polarized results within LDA.

  The spin-polarized GGA results of the present study are
summarized in Fig.~1 for the $p$(2$\times$2) geometry, where
the relevant magnitudes are plotted versus dimer buckling.
The buckling angle was imposed as a constraint in each calculation,
with the rest of the system's degrees of freedom being free
to relax.
\begin{figure}[t!]
\begin{center}
\epsfig{figure=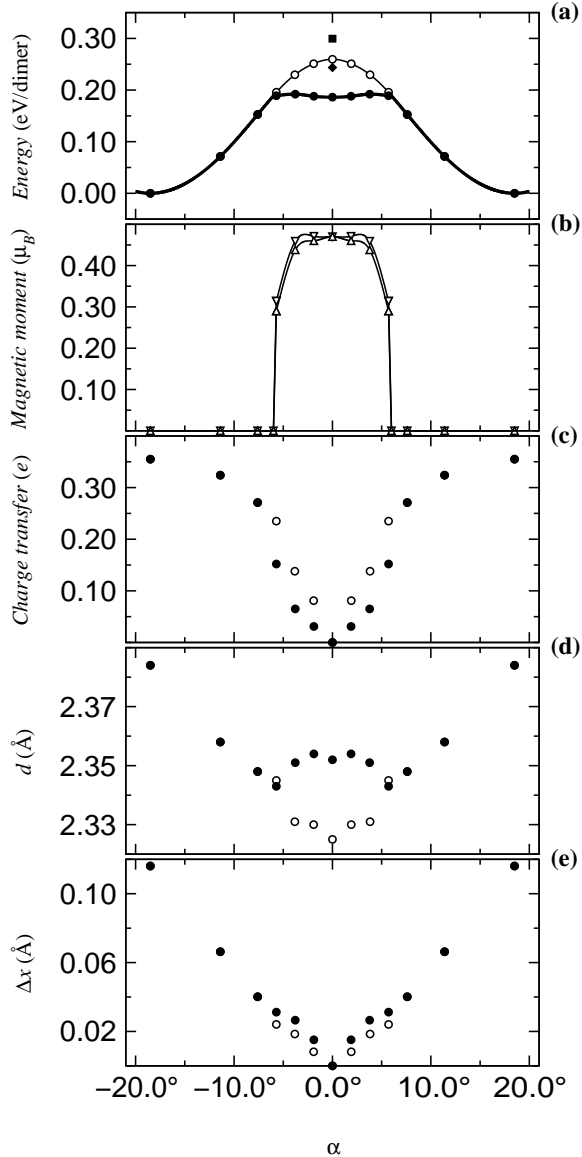,width=3in,clip=}
\end{center}
\caption{GGA results versus dimer inclination, $\alpha$, with and 
without spin polarization. The curves are interpolations.\protect\\
(a) Total energy (per dimer) with respect to the most stable $p$(2$\times$2) configuration.
The full circles are for the ground state, 
which is spin polarized (AF) at small angles; empty circles
show the unpolarized solution. The square is for the mixed AF-F 
configuration described in the text, and the diamond for the 
ferromagnetic configuration.\protect\\
(b) Magnetic moment on each dimer atom in the AF configuration.\protect\\
(c) Intradimer charge transfer ($Q_{high} - Q_{low}$), obtained from
Mulliken population analysis.\protect\\
(d) Dimer bond length.\protect\\
(e) Displacement of subsurface atoms in the direction parallel to
dimer rows.}
\label{mainfig}
\end{figure}
  As can be observed from Fig.~1, the most stable configuration
remains unpolarized, as in the LDA calculation,
with asymmetric dimers of 18.5$^{\rm o}$ of inclination. 
  At small buckling angles ($\lesssim$6.1$^{\rm o}$) the system has
a polarized, antiferromagnetic (AF) ordered solution with a lower
energy than the non-polarized solution (0.07 eV/dimer decrease at 
0$^{\rm o}$). $E(\alpha)$ seems to show a shallow minimum around the
symmetric configuration.
  The dimer bond length increases with the spin polarization.
  The shorter bond length of 2.32~\AA\ of the unpolarized solution 
can be understood in terms of the (artificial) reinforcing of the
bond (double-bond) in the dimer, essentially absent in the polarized 
solution, for which a typical Si bond length of 2.35~\AA\ is obtained.
  The subsurface lateral relaxation correlates with the 
inclination, but also with the dimer bond length.

  A magnetic moment of 0.47~$\mu_B$ is obtained at 0$^{\rm o}$
inclination for this AF configuration. For a 2$\times$1 unit cell
there are two additional spin-polarized solutions: a ferromagnetic
(F) spin configuration, and a mixed one, AF within the dimer 
and F among dimers in a row. Their energies are higher than the
AF and are shown in Fig.~1(a). The magnetic moments are 0.51~$\mu_B$
and 0.35~$\mu_B$ for the mixed and F configurations,
respectively. Interestingly, the dimer bond lengths are 
2.31~\AA\ and 2.33~\AA\ for the mixed and F configurations,
respectively, shorter than in the AF case.

  These results confirm the importance of Mott-like electronic 
correlations in this surface.
  The results do not tell, however, about the intrinsic or
extrinsic origin of the buckling that appears.
  Interdimer interactions that can make a difference are:
  ($i$) electrostatic: each dimer has an electric dipole due to
its charge transfer~\cite{Chadi79,Landemark92,Munz95} and a 
polarizability that can enhance or depress the dipole when 
interacting with its neighbors;
  ($ii$) surface stress: the relative disposition of dimers can
relax surface stress as pointed out by Garc\'{\i}a and
Northrup~\cite{alberto}; 
  ($iii$) electronic hopping: the electron delocalization
among dangling bonds of different dimers can depress electronic
correlations and change the energy.

  The first point can be discarded by noting that it has
opposite effect on the energy for the asymmetric 2$\times$1 and
$p$(2$\times$2) surfaces. 
  The difference between these two structures is in the relative
disposition of neighboring dimer inclinations: in the 2$\times$1 all
the dimers buckle in the same direction, in the $p$(2$\times$2) neighboring
dimers in a row have opposite inclinations.
  The electric dipoles are mostly parallel and repelling in the
2$\times$1 order, and antiparallel and attracting in the 
$p$(2$\times$2).
  The effect should thus appear in the energy difference 
between the asymmetric 2$\times$1 and the $p$(2$\times$2) rather
than in the difference between symmetric and asymmetric.
  This argument has been ratified 
by calculating the electrostatic interaction energy of a 
two-dimensional set of dipoles with the appropriate disposition
and the dipolar moments extracted from the DFT calculations.
  The results confirm the order
of magnitude and sign used in the argumentation: the 2$\times$1 
disposition has positive energy, whereas the $p$(2$\times$2)
has it negative, both having an order of magnitude of
0.1~eV/dimer for the maximum $p$(2$\times$2) charge transfer
of 0.36~electrons.

   It is tempting to use a similar argument for the effect of the 
surface stress: it seems reasonable to assume that by buckling the 
dimers the surface stress would behave in opposite manner for the 2$\times$1
configuration than for the $p$(2$\times$2) one. However, the calculations
do not show that trend: the compressive stress in the direction of
the dimer rows is highest for the symmetric-dimer surface, and lowest
for the $p$(2$\times$2) structure. Surface stress could thus be playing a 
role in the destabilization of the symmetric dimers.

  The effect of the electron delocalization among dimers cannot
be assessed as easily. 
  The subsistence of important correlation 
effects can be inferred from the fact that clusters with more than
one dimer, when described within quantum-chemical methods that
describe well the dynamical correlations, tend to stay
symmetric~\cite{Paulus,antonio}. However, to our knowledge,
no high-level quantum-chemical calculation comparing the buckled
against the non-buckled configurations, has been performed for
clusters with more than two surface dimers. As has been shown
recently by Penev {\it et al.}~\cite{Penev} this may be an important
point, as discussed below.
  Penev {\it et al.}~\cite{Penev}
help in discerning the issue of electron delocalization
by comparing cluster calculations
with slab calculations within the same level of theory otherwise.
  The asymmetry of the dimers is found to be very much favored in
the extended system, the clusters with one or two surface dimers
A
showing a nearly flat behavior of energy versus buckling angle,
$E(\alpha)$, close to $\alpha=0^{\rm o}$. Part of the difference, however, 
could be due to the treatment of the relaxation constraints in 
the clusters, as pointed out by Yang and Kang~\cite{Yang}.
  An approximation to the correlated {\it and} extended situation
could be obtained by
\begin{equation}
E(\alpha) \approx E^{corr}_{clus}(\alpha) + [ E^{DFT}_{slab}(\alpha) -
                                        E^{DFT}_{clus}(\alpha) ] ,
\end{equation}
where $E^{corr}_{clus}(\alpha)$ is for the correlated cluster
calculations, and $E^{DFT}_{slab}(\alpha)$ and $E^{DFT}_{clus}(\alpha)$
are for the slab and cluster DFT calculations,
respectively.
  From the calculations of Penev {\it et al.}~\cite{Penev} one can
estimate that the term
$[ E^{DFT}_{slab}(\alpha) - E^{DFT}_{clus}(\alpha) ]$ would give
a maximum gain of the order of 0.2~eV/dimer towards buckling
due to extrinsic dimer effects (comparing the slab with the
Si$_9$H$_{12}$ cluster calculation). To determine if the ground state
has a buckled or non-buckled configuration one has to know
$E^{corr}_{clus}(\alpha)$ as a function of $\alpha$ for a high-level
correlated calculation.
  A more rigorous computation for the correlated extended system
requires computation techniques not available at present. Given all
that was discussed above, a DFT calculation,
spin-polarized or not, should not be fully trusted in this respect.

  Another possibility, which can also help to gain further insight
about the correlation effects in this problem, would be the use
of a model Hamiltonian for a single row of 
dimers, as the stripe or ladders that are being treated now for
High $T_c$ superconductors, as pointed out by Kondo {\it et al.}~\cite{Kondo}.
  It would consist of a Hubbard-like model on the ladder, coupled to
a classical dynamical variable describing the buckling. Parameters
for this model can be obtained from suitable calculations with
accessible techniques, and the model can be solved in different
approximations to a high degree of accuracy. Results for it will be
presented elsewhere~\cite{future}.
  It is interesting to note that such a Hamiltonian is remarkably
close to Hamiltonians used for completely different problems in 
many-body physics. There are thus cross-breeding possibilities, and
it would be quite interesting if knowledge on the Si(100) surface
could be used to gain insights for the correlated motion of carriers
in doped perovskites.

\acknowledgements
We acknowledge useful discussions with E. A. Carter. This work has been
partially supported by the Fundaci\'on Ram\'on Areces of Spain and FAPESP
and CNPq from Brazil.


\begin{references}
\bibitem{review} 
J. A. Kubby, and J. J. Boland, Surf. Sci. Rep. {\bf 26}, 61 (1996);
H. J. W. Zandvliet, Rev. Mod. Phys. {\bf 72}, 593 (2000),
and references therein.
\bibitem{Chadi79}
D. J. Chadi, Phys. Rev. Lett. {\bf 43}, 43 (1979).
\bibitem{Artacho89}
E. Artacho, and F. Yndur\'ain, Phys. Rev. Lett. {\bf 62}, 2491 (1989).
\bibitem{Vinchon}
T. Vinchon, M. C. Desjonqu\`eres, A. M. Ole\'s, and D. Spanjaard, Phys. Rev. B
{\bf 48}, 8190 (1993).
\bibitem{Flores} 
For a review see F. Flores, J. Ortega, and R. P\'erez,
Surf. Rev. and Lett. {\bf 6}, 411 (1999).
\bibitem{Ramstad}
A. Ramstad, G. Brocks, and P. J. Kelly, Phys. Rev. B {\bf 51}, 14504 (1995),
and references therein.
\bibitem{Pauling} 
L. Pauling and Z. S. Herman, Phys. Rev. B {\bf 28}, 6154 (1983).
\bibitem{Kondo}
Y. Kondo, T. Amakusa, M. Iwatsuki, and H. Tokumoto, Surf. Sci. {\bf 453} 
L318 (2000).
\bibitem{Yokoyama}
T. Yokoyama, and K. Takayanagi, Phys. Rev. B {\bf 61}, R5078 (2000).
\bibitem{Redondo} 
A. Redondo, and W. A. Goddard III, J. Vac. Sci. Technol. {\bf 21},
344 (1982).
\bibitem{Carter}
C. J. Wu, and E. A. Carter, Chem. Phys. Lett. {\bf 185}, 172 (1991);
Phys. Rev. B {\bf 46}, 4651 (1992).
\bibitem{Paulus}
B. Paulus, Surf. Sci. {\bf 408}, 195 (1998).
\bibitem{pablo}
P. Ordej\'on, E. Artacho, and J. M. Soler, Phys. Rev. B {\bf 53},
R10441 (1996).
\bibitem{daniel}
D. S\'anchez-Portal, P. Ordej\'on, E. Artacho, and J. M. Soler,
Int. J. of Quantum Chem. {\bf 65}, 453 (1997).
\bibitem{emilio}
E. Artacho, D. S\'anchez-Portal, P. Ordej\'on, A. Garc\'{\i}a,
and J. M. Soler, Phys. Stat. Sol. (b) {\bf 215}, 809 (1999).
\bibitem{PBE} 
J. P. Perdew, K. Burke, and M. Ernzerhof,
Phys. Rev. Lett. {\bf 77}, 3865 (1996).
\bibitem{Kohn-Sham} 
W. Kohn, and L. J. Sham, Phys. Rev. {\bf 140}, 1133 (1965).
\bibitem{PZ}
J. P. Perdew, and A. Zunger, Phys. Rev. B {\bf 23}, 5075 (1981).
\bibitem{JLM} 
N. Troullier, and J. L. Martins,
Phys. Rev. B {\bf 43}, 1993 (1991).
\bibitem{KB} 
L. Kleinman, and D. M. Bylander, 
Phys. Rev. Lett. {\bf 48}, 1425 (1982).
\bibitem{Monkhorst-Pack}
H. J. Monkhorst, and J. D. Pack, Phys. Rev. B {\bf 13}, 5188 (1976).
\bibitem{Moreno}
J. Moreno, and J. M. Soler, Phys. Rev. B {\bf 45}, 13891 (1992).
\bibitem{Exp.values}
{\it Numerical Data and Functional Relationships in Science and Technology},
edited by K. H. Hellwege and O. Madelung, Landolt-B\"ornstein, New Series,
Group III, Vol. 17a (Springer, Berlin, 1982).
\bibitem{PW.values}
A. Garc\'{\i}a, C. Els\"asser, J. Zhu, S. G. Louie, and M. L. Cohen, Phys.
Rev. B {\bf 46}, 9829 (1992).
\bibitem{javier}
A PW calculation with the pseudopotential used in this work
and a 50 Ry cutoff yields a lattice parameter of 5.38~\AA\ and a 
bulk modulus of 96~GPa for the Murnaghan fit, and 98~GPa for the quartic 
fit~\cite{junquera}.
\bibitem{junquera} J. Junquera, unpublished.
\bibitem{murnaghan}
F. D. Murnaghan, Proc. Natl. Acad. Sci. U.S.A. {\bf 30}, 244 (1944).
\bibitem{surf. energy}
The surface energy per 1$\times$1 unit cell for the $p$(2$\times$2) configuration
is defined as
\mbox{$E_s=\frac{1}{4}(E_T-n_{Si}E_{Si}^{bulk}-n_HE_H)$.} $E_T$ is the total
energy, $E_{Si}^{bulk}$ is the energy
per bulk silicon atom, and $E_H$ is the energy per saturating hydrogen atom
(obtained from
the Si-H bond relaxation calculation). $n_{Si}$ and $n_H$ are
the number of
silicon and hydrogen atoms in the cell respectively.
\bibitem{Dabrowski92}
J. D\c{a}browski, and M. Scheffler, Appl. Surf. Sci {\bf 56-58}, 15 (1992).
\bibitem{Fritsch95}
J. Fritsch, and P. Pavone, Surf. Sci. {\bf 344}, 159 (1995).
\bibitem{Kruger93}
P. Kr\"uger, and J. Pollmann, Phys. Rev. B {\bf 47}, 1898 (1993).
\bibitem{alberto}
A. Garc\'{\i}a, and J. E. Northrup, Phys. Rev. B {\bf 48}, 17350 (1993).
\bibitem{Northrup93}
J. E. Northrup, Phys. Rev. B {\bf 47}, 10032 (1993).
\bibitem{Landemark92}
E. Landemark, C. J. Karlsson, Y.-C. Chao, and R. I. G. Uhrberg,
Phys. Rev. Lett. {\bf 69}, 1588 (1992).
\bibitem{Munz95}
A. W. Munz, Ch. Ziegler, and W. G\"opel, Phys. Rev. Lett. {\bf 74}, 2244 (1995).
\bibitem{antonio}
A. J. R. da Silva, unpublished.
\bibitem{Penev}
E. Penev, P. Kratzer, and M. Scheffler, J. Chem. Phys. {\bf 110}, 3986
(1999).
\bibitem{Yang}
C. Yang, and H. Chuan Kang, J. Chem. Phys. {\bf 110}, 11029 (1999).
\bibitem{future} 
\'O. Paz, A. J. R. da Silva, and E. Artacho, in preparation.
\end{references}
\end{document}